\documentclass[reqno,11ptt]{amsart}
\usepackage{amsmath,amsthm}
\usepackage{amsfonts,amssymb}

\usepackage{enumerate}

\theoremstyle{definition}

\numberwithin{equation}{section}

\newcommand{\Rset}{\mathbb{R}}

\newcommand{\Zset}{\mathbb{Z}}
\newcommand{\Nset}{\mathbb{N}}

\newcommand{\Sset}{\mathbb{S}}
\newcommand{\Tset}{\mathbb{T}}

\newcommand{\Id}{\mathrm{Id}}
\newcommand{\Span}{\mathrm{span}}

\newcommand{\al}{{\alpha}}
\newcommand{\be}{{\beta}}
\newcommand{\la}{{\lambda}}

\newcommand{\gat}{{\tilde{\gamma}}}

\newcommand{\ga}{{\gamma}}

\newcommand{\pd}{\partial}

 \newcommand{\bv}{{\mathbf v}}
 \newcommand{\bx}{{\mathbf x}}
 
\newcommand{\bz}{{\mathbf z}}

\newcommand{\bga}{\boldsymbol{\gamma}}
\newcommand{\bnu}{\boldsymbol{\nu}}

\newcommand{\bbe}{\boldsymbol{\beta}}

\newcommand{\cG}{{\mathcal G}}
\newcommand{\cL}{{\mathcal L}}

\newcommand{\cM}{{\mathcal M}}

\newcommand{\cH}{{\mathcal H}}
\newcommand{\cB}{{\mathcal B}}

\newcommand{\cV}{{\mathcal V}}
\newcommand{\cW}{{\mathcal W}}

\newcommand{\cLh}{\hat{{\mathcal L}}}

\newcommand{\fA}{{\mathfrak A}}

\newcommand{\cDz}{{\mathcal D}_z}

\begin{document}

\title[Symmetry algebra for the generic system on the sphere]{Symmetry algebra for the generic superintegrable system on the sphere}

\date{January 25, 2018}

\author{Plamen Iliev}
\address{School of Mathematics, Georgia Institute of Technology, Atlanta, GA 30332--0160, USA}
\email{iliev@math.gatech.edu}
\thanks{The author is partially supported by Simons Foundation Grant \#280940.}

\begin{abstract}
The goal of the present paper is to provide a detailed study of irreducible representations of the algebra generated by the symmetries of the generic quantum superintegrable system on the $d$-sphere. Appropriately normalized, the symmetry operators preserve the space of polynomials. Under mild conditions on the free parameters, maximal abelian subalgebras of the symmetry algebra, generated by Jucys-Murphy elements, have unique common eigenfunctions consisting of  families of Jacobi polynomials in $d$ variables. We describe the action of the symmetries on the basis of Jacobi polynomials in terms of multivariable Racah operators, and combine this with different embeddings of symmetry algebras of lower dimensions to prove that the representations restricted on the space of polynomials of a fixed total degree are irreducible.
\end{abstract}
\maketitle

\section{Introduction}\label{se1}

Recall that a quantum superintegrable system is an integrable Hamiltonian system on a $d$-dimensional Riemannian or pseudo-Riemannian manifold with potential: 
$$\cH = \Delta + V$$ 
that admits the maximum possible $2d-1$ algebraically independent partial differential operators $\cH_k$ commuting with $\cH$, i.e. 
$$[\cH,\cH_k]=0.$$
These systems appear in a wide variety of modern physical and mathematical theories, from semiconductors to black holes. For a thorough account of the general theory and its numerous applications we refer the reader to the review article by Miller, Post and Winternitz \cite{MPW}. 

Many important examples of superintegrable systems can be obtained through limits from the so called generic quantum superintegrable system on the sphere, with potential 
$$V(y)=\sum_{k=1}^{d+1}\frac{b_k}{y_k^2},$$
where $y$ belongs to the $d$-sphere, and  $\{b_k\}_{k=1,\dots,d+1}$ are free parameters. The representation theory of the algebra generated by the symmetries of this system  has attracted a lot of attention recently and it turned out to be closely related to multivariate extensions of the Askey scheme of hypergeometric orthogonal polynomials and their bispectral properties, the Racah problem for $\mathfrak{su}(1,1)$, representations of the Kohno-Drinfeld algebra, the Laplace-Dunkl operator associated with $\Zset_2^{d+1}$ root system; see for instance \cite{DGVV,GV,GVZ,I,KMP1,KMP2,MT,Post} and the references therein. 

The goal of the present paper is two fold. First, we describe the representations of the symmetry algebra for the generic quantum superintegrable system on the sphere in terms of the multivariable Racah operators introduced in \cite{GI} on the space of polynomials in several variables. We provide a detailed account of the theory, building and expanding on the recent work \cite{I}.  In dimensions $2$ and $3$, the approach is based on first principles which naturally leads to precise constraints on the free parameters $\{b_k\}$ for which these constructions can be applied. In particular, it explains how these constraints fix uniquely the underlying basis of Jacobi polynomials as eigenfunctions of maximal abelian subalgebras generated by Jucys-Murphy elements. The second goal of the paper is to combine these results and constructions with different embeddings of the symmetry algebras of lower dimensions to prove that the representations are irreducible. 

While the arguments apply in arbitrary dimension, the difficulty increases significantly as the dimension grows, and dimension $d\geq 4$ requires conceptually new ingredients.  For that reason, we treat separately the $2$-dimensional and the $3$-dimensional sphere, where we can write simple closed formulas for all symmetry operators. The explicit formulas here serve as important building blocks for the constructions in higher dimensions, and provide an alternative approach to several earlier works. In dimension $d\geq 4$, the picture changes drastically - linear combinations of the Racah operators are no longer sufficient to describe the representations of the symmetry algebra. To see this, we discuss the linear independence of the second-order symmetries, which combined with the total number of available Racah operators explains the need to search for algebraic generators.  Thus, we arrive at a fourth-order relation which allows to construct an explicit set consisting of $2d-1$ algebraic generators for the symmetry algebra. 

The paper is organized as follows. In the next section, we introduce the generic quantum superintegrable system on the sphere, its symmetry algebra and an appropriate gauge transformation, which allows to induce representations on the space of polynomials of several variables. In Section~\ref{se3} and \ref{se4}, we describe the representations of the symmetry algebra for the $2$-sphere and the $3$-sphere, respectively, and we prove that these representations are irreducible. In Section \ref{se5}, we outline the constructions in arbitrary dimension together with a detailed proof of the irreducibility.

\section{The model and its normalization}\label{se2}

We denote by $\Sset^{d}=\{y\in\Rset^{d+1}:y_1^2+\cdots+y_{d+1}^2=1\}$ the $d$-dimensional sphere in $\Rset^{d+1}$ and we set $\pd_{y_j}=\frac{\pd}{\pd {y_j}}$ for the partial derivative with respect to $y_j$, for $j=1,\dots,d+1$. 
The  generic superintegrable system on the sphere is the quantum system with Hamiltonian
\begin{equation}\label{2.1}
\cH=\Delta_{\Sset^{d}}+\sum_{k=1}^{d+1}\frac{b_k}{y_k^2},
\end{equation}
where 
$$\Delta_{\Sset^{d}}=\sum_{1\leq i<j \leq d+1}\left(y_i{\pd_{y_j}}-y_j{\pd_{y_i}}\right)^2$$
is the Laplace-Beltrami operator on the sphere and $\{b_k\}_{k=1,\dots,d+1}$ are free parameters.  If we set
\begin{equation}\label{2.2}
\cH_{i,j}=\left(y_i{\pd_{y_j}}-y_j{\pd_{y_i}}\right)^2+\frac{b_iy_j^2}{y_i^2}+\frac{b_jy_i^2}{y_j^2}
\end{equation}
then it is straightforward to check that 
$$[\cH,\cH_{i,j}]=0$$
and therefore the operators $\cH_{i,j}$ represent symmetries, or integrals of motion for $\cH$. The Hamiltonian can be decomposed in terms of the symmetry operators as follows
$$\cH=\sum_{1\leq i<j\leq d+1}\cH_{i,j}+\sum_{k=1}^{d+1}b_k.$$
Next, we pick appropriate coordinates and a gauge factor, so that the symmetry operators preserve the space of polynomials. Let 
\begin{equation}\label{2.3}
x_i=y_i^2 \qquad \text{ and }\qquad b_i=\frac{1}{4}-\ga_i^2, \qquad\text{ for }i=1,\dots,d+1.
\end{equation}
In the remaining part of the paper, we will work with the variables $x_i$ and the parameters $\ga_i$, related to the original variables and parameters by \eqref{2.3}. In particular, we will impose later mild restrictions on the parameters $\ga_i$, which can easily be translated onto the parameters $b_i$ using the connection in \eqref{2.3}.
If $\cG(x)$ denotes the gauge factor
\begin{equation*}
\cG(x)=\prod_{j=1}^{d+1}x_j^{-\frac{\ga_j}{2}-\frac{1}{4}},
\end{equation*}
then it is straightforward to check that
\begin{equation}\label{2.4}
\begin{split}
&\frac{1}{4}\cG(x)\cH_{i,j}\, \cG^{-1}(x)+\frac{(\ga_i+1)(\ga_j+1)}{2}-\frac{1}{8}\\
&\qquad= x_ix_j\left({\pd_{x_j}}-{\pd_{x_i}}\right)^2+\left[(\ga_i+1)x_j-(\ga_j+1)x_i\right] \left({\pd_{x_i}}-{\pd_{x_j}}\right).
\end{split}
\end{equation}
For a vector $z\in\Rset^s$ we denote by 
$$|z|=z_1+\cdots+z_{s}$$ 
the sum of its coordinates. If we fix  
$x=(x_1,\dots,x_d)$ as coordinates, then the fact that $y\in\Sset^d$ implies that $x_{d+1}=1-|x|$. 
If we denote by $\cL_{i,j}$ the operator on the right-hand side of equation~\eqref{2.4}, 
we obtain the following expression in the coordinates $x=(x_1,\dots,x_d)$:
\begin{align}
&\cL_{i,j}=x_ix_j(\pd_{x_i}-\pd_{x_j})^2+[(\ga_i+1)x_j-(\ga_j+1)x_i](\pd_{x_i}-\pd_{x_j}), \nonumber\\
&\hskip7cm  \text{ if }i\neq j\in\{1,\dots,d\},\label{2.5}\\
\intertext{and}
&\cL_{j,d+1}=\cL_{d+1,j}=x_j(1-|x|)\pd_{x_j}^2+[(\ga_j+1)(1-|x|)-(\ga_{d+1}+1)x_j]\pd_{x_j}, \nonumber\\
&\hskip7cm  \text{ if }j\in\{1,\dots,d\}.\label{2.6}
\end{align}
With the above notations we also have
\begin{equation}\label{2.7}
\frac{1}{4}\cG(x)\cH\, \cG^{-1}(x)=\cL+\mathrm{constant},
\end{equation}
where
\begin{equation}\label{2.8}
\begin{split}
\cL&=\sum_{1\leq i<j\leq d+1}\cL_{i,j}\\
&=\sum_{k=1}^{d}x_k(1-x_k)\pd^2_{x_k} 
 -2\sum_{1\leq k<j\leq d} x_kx_j\pd_{x_k}\pd_{x_j}+\sum_{k=1}^{d}(\ga_k+1-(|\ga|+d+1) x_k)\pd_{x_k},
\end{split}
\end{equation}
and $|\ga|=\ga_1+\cdots+\ga_{d+1}$.
The above computations show that we can replace the Hamiltonian $\cH$ and its symmetry operators $\cH_{i,j}$ by the operator $\cL$ and its symmetry operators $\cL_{i,j}$. Note that the operators $\cL_{i,j}$ have polynomial coefficients and therefore they preserve the space of polynomials of $x_1,\dots,x_d$. In the next sections, we use this fact to construct irreducible representations of the algebra $\fA_d(\ga)$ generated by the operators $\cL_{i,j}$, which in view of the above computations, corresponds to the symmetry algebra for the generic quantum superintegrable system on the sphere $\Sset^d$. Therefore, we refer to $\fA_d(\ga)$ as the {\em symmetry algebra} for the generic system on the sphere.

The operator $\cL$ defined in \eqref{2.8} has a long history in the mathematical literature. It first appeared in the monograph \cite{AK} in the case $d = 2$ in connection with differential equations satisfied by the Lauricella functions. Its link to the superintegrable system on the sphere 
in arbitrary dimension was revealed in \cite{KMT}.  

Finally, we note yet another link which plays an important role in the constructions. From the explicit formulas in equation \eqref{2.5}-\eqref{2.6} it is easy to see that the operators $\cL_{i,j}$ satisfy the following commutativity relations:
\begin{align}
&[\cL_{i,j},\cL_{k,l}]=0, &\text{ if }i,j,k,l \text{ are distinct,}\label{2.9}\\
&[\cL_{i,j},\cL_{i,k}+\cL_{j,k}]=0, &\text{ if }i,j,k \text{ are distinct.} \label{2.10}
\end{align}
These relations show that the symmetry operators provide a representation of the Kohno-Drinfeld algebra which appears in the structure of the holonomy of the Knizhnik--Zamolodchikov connection and the representation of the braid group, see for instance \cite{Ko}.

\section{Symmetry algebra for the $2$-sphere}\label{se3}
\subsection{Construction of the module}\label{ss3.1}
When $d=2$, the symmetry algebra is generated by the $3$ operators: $\cL_{1,2}$, $\cL_{1,3}$, $\cL_{2,3}$ which commute with $\cL=\cL_{1,2}+\cL_{1,3}+\cL_{2,3}$. The operators $\cL$ and $\cL_{2,3}$  can be simultaneously diagonalized on the space of polynomials and the eigenfunctions can be written explicitly in terms of appropriate two-variable Jacobi polynomials. This was discovered by Proriol \cite{Prorial}, building on the work \cite{MP} where an important particular case appeared in the study of the Schr\"odinger equation for helium. To describe this construction, we define the classical Jacobi polynomial of degree $n$ with parameters $\al$ and $\be$ via the formula
\begin{equation*}
p_n^{(\al,\be)}(t) =  \frac{(\al+1)_n}{(\be+1)_n} \sum_{k=0}^{n}\frac{(-n)_k(n+\al+\be+1)_k}{k!(\al+1)_k}\left(\frac{1-t}{2}\right)^k,
\end{equation*}
where $(a)_k$  denotes the Pochhammer symbol:
$$(a)_0=1\qquad \text{ and }\qquad (a)_k=a(a+1)\cdots(a+k-1)\text{ for } k\in\Nset.$$
We use $\Zset_{\leq k}$ to denote the set of all integers less than or equal to $k$. 
Note that the Jacobi polynomial $p_n^{(\al,\be)}(t) $ is a well-defined polynomial of degree $n$ when the parameters $\al$ and $\be$ satisfy the following conditions
\begin{equation}\label{3.1}
\al\notin \Zset_{\leq -1}, \qquad \be\notin \Zset_{\leq -1}, \qquad \al+\be\notin \Zset_{\leq -2}.
\end{equation}
These conditions are automatically satisfied when $\al>-1$ and $\be>-1$. In this case, the polynomials $p_n^{(\al,\be)}(t)$ are mutually orthogonal with respect to the weight $(1-t)^{\al}(1+t)^{\be}$ on $(-1,1)$.

Next, we define polynomials in two variables $x=(x_1,x_2)$, with degree indices $\nu_1$ and $\nu_2$, depending on the three parameters $\ga=(\ga_1,\ga_2,\ga_3)$ in terms of the one-variable Jacobi polynomials as
\begin{equation} \label{3.2}
P_{\nu_1,\nu_2}(x;\ga) = p_{\nu_1}^{(\ga_2+\ga_3+2\nu_2+1,\ga_1)}(2x_1-1) (1-x_1)^{\nu_2} p_{\nu_2}^{(\ga_3,\ga_2)}\left (\frac{2 x_2}{1-x_1}-1\right).
\end{equation} 
These polynomials are well-defined if the parameters $\ga_1,\ga_2,\ga_3$ satisfy the following conditions
\begin{equation}\label{3.3}
\ga_1\notin \Zset_{\leq -1}, \; \ga_2\notin \Zset_{\leq -1}, \; \ga_3\notin \Zset_{\leq -1}, \; \ga_2+\ga_3\notin \Zset_{\leq -2}, \;\ga_1+ \ga_2+\ga_3\notin \Zset_{\leq -3}.
\end{equation}
The polynomials $P_{\nu_1,\nu_2}(x;\ga)$ are eigenfunctions of the commuting operators $\cL$ and $\cL_{2,3}$ and 
satisfy the spectral equations
\begin{align}
\cL P_{\nu_1,\nu_2} (x;\ga)&=-(\nu_1+\nu_2)(\nu_1+\nu_2+\ga_1+\ga_2+\ga_3+2)P_{\nu_1,\nu_2} (x;\ga),\label{3.4}\\
\cL_{2,3} P_{\nu_1,\nu_2} (x;\ga)&=-\nu_2(\nu_2+\ga_2+\ga_3+1)P_{\nu_1,\nu_2} (x;\ga).\label{3.5}
\end{align}
For $n\in\Nset_0$ we denote by $\cV^{2}_n(\ga)$ the space spanned by the two-variable polynomials in \eqref{3.2} of total degree $n$, i.e. we set
\begin{equation}\label{3.6}
\cV^{2}_n(\ga)=\Span\{P_{\nu_1,\nu_2}(x;\ga):\nu_1+\nu_2=n\}.
\end{equation}
We already know that the operators $\cL$ and $\cL_{2,3}$ act diagonally on the basis $P_{\nu_1,\nu_2}$. It turns out that the operators $\cL_{1,2}$ and $\cL_{1,3}$ preserve this space and therefore $\cV^{2}_n(\ga)$ is a module of the symmetry algebra $\fA_2(\ga)$ generated by the operators $\cL_{1,2}$, $\cL_{1,3}$ and $\cL_{2,3}$. To describe their action, we denote by $\Id$ the identity operator and we define shift operators $E_{\nu_j}$ acting on functions of $\nu=(\nu_1,\nu_2)$ by
$$E_{\nu_1}f_{\nu_1,\nu_2}=f_{\nu_1+1,\nu_2}\qquad \text{ and }\qquad E_{\nu_2}f_{\nu_1,\nu_2}=f_{\nu_1,\nu_2+1}.$$
With these notations, we consider the recurrence operator acting on functions of $\nu=(\nu_1,\nu_2)$ by
\begin{equation} \label{3.7}
\cB_{1,2}=c^{-,+}_{\nu_1,\nu_2}E_{\nu_1}^{-1}E_{\nu_2}+c^{0,0}_{\nu_1,\nu_2}\Id + c^{+,-}_{\nu_1,\nu_2}E_{\nu_1}E_{\nu_2}^{-1},
\end{equation} 
with coefficients
\begin{align*}
c^{-,+}_{\nu_1,\nu_2}&=\frac{\nu_1(\ga_2 + \nu_2 +1) ( \ga_2 + \ga_3 + \nu_2 +1) ( \ga_1 + \ga_2 + \ga_3 + \nu_1 + 2 \nu_2 +2)}{( \ga_2 + \ga_3 + 2 \nu_2 +1) (\ga_2 + \ga_3 + 2 \nu_2 + 2)},\\
c^{0,0}_{\nu_1,\nu_2}&= -(\nu_1 + \nu_2 + 2 \nu_1 \nu_2 + \nu_2 \ga_1\ + \nu_1 \ga_2)\\
&\quad+\frac{(\nu_1+1) \nu_2 (\ga_1 + \nu_1 + 1) ( \ga_2+\nu_2)}{ \ga_2 + \ga_3+2 \nu_2}-\frac{\nu_1 (\nu_2+1) ( \ga_1+\nu_1) ( \ga_2 + \nu_2 +1)}{ \ga_2 + \ga_3 + 2 \nu_2 +2},\\
c^{+,-}_{\nu_1,\nu_2}&=\frac{\nu_2 (\ga_1 + \nu_1 +1) (\ga_3 + \nu_2 ) (\ga_2 + \ga_3 + \nu_1 + 2 \nu_2 + 1)}{(\ga_2 + \ga_3+2 \nu_2 ) (\ga_2 + \ga_3 + 2 \nu_2 +1)}.
\end{align*}
Note that for $n\in\Nset_0$ we can restrict its action on functions defined for $\nu_1+\nu_2=n$ since both operators $E_{\nu_1}^{-1}E_{\nu_2}$ and $E_{\nu_1}E_{\nu_2}^{-1}$ preserve this condition.
Using the explicit formulas above, one can show that this operator represents the action of $\cL_{1,2}$ on the basis $P_{\nu_1,\nu_2}(x;\ga)$ of $\cV^{2}_n(\ga)$, i.e. we have
\begin{equation} \label{3.8}
\cL_{1,2} P_{\nu_1,\nu_2}(x;\ga) =\cB_{1,2} P_{\nu_1,\nu_2}(x;\ga).
\end{equation} 
This shows that the space $\cV^{2}_n(\ga)$ is a module over the symmetry algebra $\fA_2(\ga)$. The explicit action of $\cL_{1,3}$ on the basis $P_{\nu_1,\nu_2}(x;\ga)$ can be easily deduced by writing $\cL_{1,3}$ as 
$$\cL_{1,3}=\cL-\cL_{1,2}-\cL_{2,3},$$
and using equations \eqref{3.4}, \eqref{3.5} and \eqref{3.8}.

We note that, appropriately normalized, the operator $\cB_{1,2}$ corresponds to the recurrence operator for the Racah polynomials. This fact was discovered in the work of Kalnins, Miller and Post \cite{KMP1} in different notations, where the representations of the symmetry algebra for the $2$-sphere were investigated. A different interpretation, related to representations of $\mathrm{su}(1,1)$ was obtained by Genest, Vinet and Zhedanov \cite{GVZ}. We outline yet another explanation based on orthogonal polynomials, which is close to the presentation above, and which was used in \cite{I} to obtain explicit formulas in arbitrary dimension in terms of the Racah operators introduced in \cite{GI}.  If $\ga_j>-1$ for $j=1,2,3$, then the polynomials $P_{\nu_1,\nu_2}(x;\ga)$ in \eqref{3.2} are mutually orthogonal with respect to the weight 
$$w_2(x_1,x_2)=x_1^{\ga_1}x_2^{\ga_2}(1-x_1-x_2)^{\ga_3}$$
on the simplex $\Tset^2=\{(x_1,x_2)\in\Rset^2:0\leq x_1,\,0\leq x_2,\, x_1+x_2\leq 1\}$. Moreover, one can show that the operators $\cL_{1,2}$, $\cL_{1,3}$, $\cL_{2,3}$ are self-adjoint with respect to the inner product induced by $w_2$. This means that the symmetry algebra will preserve the space $\cV^{2}_n(\ga)$ of orthogonal polynomials of fixed total degree $n$. Note also that $w_2$ is invariant if we permute simultaneously $(\ga_1,\ga_2,\ga_3)$ and $(x_1,x_2,x_3)$, where $x_3=1-x_1-x_2$. This shows that we can construct other bases of $\cV^{2}_n(\ga)$ by applying permutations.  The transition matrices between these different orthogonal bases can be expressed in terms of the Racah weight and polynomials as shown by Dunkl in \cite{Du84}. In particular, the result of Dunkl which connects $P_{\nu_1,\nu_2}(x;\ga)$ to the basis on which $\cL_{1,2}$ acts diagonally can be used to express $\cB_{1,2}$ in terms of the Racah operator.

\subsection{Irreducibility}\label{ss3.2}
We show below that $\cV^{2}_n(\ga)$ is an irreducible module over $\fA_2(\ga)$. Let us denote by 
$$\la_1(\nu_1,\nu_2)=-(\nu_1+\nu_2)(\nu_1+\nu_2+\ga_1+\ga_2+\ga_3+2)\text{ and }\la_2(\nu_2)=-\nu_2(\nu_2+\ga_2+\ga_3+1)$$ 
the eigenvalues in equations \eqref{3.4}-\eqref{3.5}.
Note that if $\nu_2,\mu_2\in\Nset_0$ then 
$$\la_2(\nu_2)-\la_2(\mu_2)=-(\nu_2-\mu_2)(\nu_2+\mu_2+\ga_2+\ga_3+1)\neq 0, \text{ if }\nu_2\neq \mu_2,$$
since the second term cannot be zero by \eqref{3.3}. A similar computation shows that 
$$\la_1(\nu_1,\nu_2)-\la_1(\mu_1,\mu_2)\neq 0,\text{ if }\nu_1+\nu_2\neq \mu_1+\mu_2.$$
Therefore, if $\nu=(\nu_1,\nu_2)\in\Nset_0^2$ and $\mu=(\mu_1,\mu_2)\in\Nset_0^2$ then 
$$(\la_1(\nu_1,\nu_2),\la_2(\nu_2))= (\la_1(\mu_1,\mu_2),\la_2(\mu_2))\text{ if and only if }(\nu_1,\nu_2)= (\mu_1,\mu_2).$$
This shows that the spectral equations \eqref{3.4}-\eqref{3.5} characterize uniquely the polynomials $P_{\nu_1,\nu_2}(x;\ga)$ in the space of polynomials, up to unessential multiplicative constants. 

Fix now $n\in\Nset$ and suppose that $\cV\neq \{0\}$ is a nontrivial $\fA_2(\ga)$-submodule of $\cV^{2}_n(\ga)$. Since $\cL_{2,3}$ has distinct eigenvalues on $\cV^{2}_n(\ga)$, it follows that $\cL_{2,3}$ can be diagonalized on $\cV$. Therefore, there exists at least one polynomial $P_{\nu_1,\nu_2}$ which belongs to $\cV$. We want to show next that
\begin{itemize}
\item[(i)] If $P_{\nu_1,\nu_2}(x;\ga)\in\cV$ and $\nu_1>0$ then  $P_{\nu_1-1,\nu_2+1}(x;\ga)\in\cV$;
\item[(ii)] If $P_{\nu_1,\nu_2}(x;\ga)\in\cV$ and $\nu_2>0$ then  $P_{\nu_1+1,\nu_2-1}(x;\ga)\in\cV$.
\end{itemize}
Indeed, if (i) and (ii) hold then it is easy to see that $\cV$ must contain all polynomials $P_{\nu_1,\nu_2}$ of total degree $n$, hence $\cV=\cV^{2}_n(\ga)$.

Note first that
\begin{equation}\label{3.9}
c^{-,+}_{\nu_1,\nu_2}\neq 0, \qquad \text{ when }\nu_1>0.
\end{equation}
Therefore, if $\nu_2=0$, the statement in (i) follows from the fact that 
$$\cL_{1,2} P_{\nu_1,0} (x;\ga)-c^{0,0}_{\nu_1,0}P_{\nu_1,0} (x;\ga)=c^{-,+}_{\nu_1,0}P_{\nu_1-1,1} (x;\ga)\in \cV.$$
Similarly, we see that 
\begin{equation}\label{3.10}
c^{+,-}_{\nu_1,\nu_2}\neq 0, \qquad \text{ when }\nu_2>0,
\end{equation}
which implies that (ii) holds if $\nu_1=0$. It remains to show that (i) and (ii) hold when both $\nu_1$ and $\nu_2$ are positive. Since $\cL_{2,3}$ is a differential operator acting on the variables $x_1$, $x_2$, and $\cB_{1,2}$ is a difference operator acting on the indices $\nu_1,\nu_2$, it follows that 
$$[\cL_{2,3},\cB_{1,2}]P_{\nu_1,\nu_2}(x;\ga)=0.$$
Combining this with equations \eqref{3.5} and \eqref{3.8}, we compute the action of the commutator of the operators $\cL_{2,3}$ and $\cL_{1,2}$ on $P_{\nu_1,\nu_2}$:
\begin{align*}
[\cL_{2,3},\cL_{1,2}]P_{\nu_1,\nu_2}(x;\ga)&=(\cL_{2,3}\cL_{1,2}-\cL_{1,2}\cL_{2,3})P_{\nu_1,\nu_2}(x;\ga)\\
&=(\cL_{2,3}\cB_{1,2}-\cL_{1,2}\la_2(\nu_2))P_{\nu_1,\nu_2}(x;\ga)\\
&=(\cB_{1,2}\cL_{2,3}-\la_2(\nu_2)\cL_{1,2})P_{\nu_1,\nu_2}(x;\ga)\\
&=(\cB_{1,2}\la_2(\nu_2)-\la_2(\nu_2)\cB_{1,2})P_{\nu_1,\nu_2}(x;\ga)\\
&=[\cB_{1,2},\la_2(\nu_2)]P_{\nu_1,\nu_2}(x;\ga)\\
&=(d^{-,+}_{\nu_1,\nu_2}E_{\nu_1}^{-1}E_{\nu_2}+ d^{+,-}_{\nu_1,\nu_2}E_{\nu_1}E_{\nu_2}^{-1})P_{\nu_1,\nu_2}(x;\ga),
\end{align*}
where 
\begin{align}
d^{-,+}_{\nu_1,\nu_2}&=(\la_2(\nu_2+1)-\la_2(\nu_2))c^{-,+}_{\nu_1,\nu_2},\label{3.11}\\
d^{+,-}_{\nu_1,\nu_2}&=(\la_2(\nu_2-1)-\la_2(\nu_2))c^{+,-}_{\nu_1,\nu_2}.\label{3.12}
\end{align}
Since 
\begin{align*}
&\cL_{1,2} P_{\nu_1,\nu_2} (x;\ga)-c^{0,0}_{\nu_1,\nu_2}P_{\nu_1,\nu_2} (x;\ga)\\
&\qquad =c^{-,+}_{\nu_1,\nu_2}P_{\nu_1-1,\nu_2+1} (x;\ga)+ c^{+,-}_{\nu_1,\nu_2}P_{\nu_1+1,\nu_2-1} (x;\ga)\in \cV,\\
\intertext{and}
&[\cL_{2,3},\cL_{1,2}]P_{\nu_1,\nu_2}(x;\ga)\\
&\qquad =d^{-,+}_{\nu_1,\nu_2}P_{\nu_1-1,\nu_2+1} (x;\ga)+ d^{+,-}_{\nu_1,\nu_2}P_{\nu_1+1,\nu_2-1} (x;\ga)\in \cV,
\end{align*}
we can deduce that $P_{\nu_1+1,\nu_2-1} (x;\ga)\in \cV$ and $P_{\nu_1-1,\nu_2+1} (x;\ga)\in \cV$, if we can show that 
\begin{equation*}
D=\det\left[\begin{matrix} c^{-,+}_{\nu_1,\nu_2} &  c^{+,-}_{\nu_1,\nu_2}\\
d^{-,+}_{\nu_1,\nu_2} &  d^{+,-}_{\nu_1,\nu_2}  \end{matrix}\right]\neq 0.
\end{equation*}
Using equations \eqref{3.11} and \eqref{3.12} we see that
\begin{equation*}
D=2c^{-,+}_{\nu_1,\nu_2}  c^{+,-}_{\nu_1,\nu_2}(1 + 2\nu_2 + \ga_2 + \ga_3).
\end{equation*}
Equations \eqref{3.9}, \eqref{3.10} and \eqref{3.3} now imply that $D\neq 0$, completing the proof of the irreducibility. 

\section{Symmetry algebra for the $3$-sphere}\label{se4}
\subsection{Construction of the module}\label{ss4.1}
When $d=3$, we have $6$ second-order symmetry operators $\cL_{1,2}$, $\cL_{1,3}$, $\cL_{1,4}$, $\cL_{2,3}$, $\cL_{2,4}$, $\cL_{3,4}$ of the operator 
$$\cL=\cL_{1,2}+\cL_{1,3}+\cL_{1,4}+\cL_{2,3}+\cL_{2,4}+\cL_{3,4}.$$
To simplify the notations, we set $\ga_{ij}=\ga_{i}+\ga_j$, $\nu_{ij}=\nu_i+\nu_j$ for distinct indices $i$ and $j$, and similarly $\ga_{ijk}=\ga_i+\ga_j+\ga_k$, $\ga_{1234}=\ga_1+\ga_2+\ga_3+\ga_4$, etc. For distinct indices $i,j,k$ we also set 
\begin{equation}\label{4.1}
\cL_{i,j,k}=\cL_{i,j}+\cL_{i,k}+\cL_{j,k}.
\end{equation}
The operators $\cL_{3,4}$, $\cL_{2,3,4}$ and $\cL$ commute and can be simultaneously diagonalized by appropriate Jacobi polynomials in $3$ variables as follows. 
We define polynomials in $x=(x_1,x_2,x_3)$, with degree indices $\nu=(\nu_1,\nu_2,\nu_3)$, depending on the four parameters $\ga=(\ga_1,\ga_2,\ga_3,\ga_4)$ in terms of the one-variable Jacobi polynomials as
\begin{equation} \label{4.2}
\begin{split}
P_{\nu}(x;\ga) &= p_{\nu_1}^{(\ga_{234}+2\nu_{23}+2,\ga_1)}(2x_1-1) (1-x_1)^{\nu_2} p_{\nu_2}^{(\ga_{34}+2\nu_3+1,\ga_2)}\left (\frac{2 x_2}{1-x_1}-1\right)\\
&\qquad \times  (1-x_1-x_2)^{\nu_3} p_{\nu_3}^{(\ga_4,\ga_3)}\left (\frac{2 x_3}{1-x_1-x_2}-1\right).
\end{split}
\end{equation} 
These polynomials are well-defined if the parameters satisfy the following conditions
\begin{equation}\label{4.3}
\begin{split}
\ga_j\notin \Zset_{\leq -1}, \text{ for }j=1,2,3,4,\quad  \ga_{34}\notin \Zset_{\leq -2},  \quad
  \ga_{234}\notin \Zset_{\leq -3},\quad \ga_{1234}\notin \Zset_{\leq -4}.
\end{split}
\end{equation}
The polynomials $P_{\nu}(x;\ga)$ are eigenfunctions of the commuting operators $\cL$, $\cL_{2,3,4}$, $\cL_{3,4}$ and 
satisfy the following spectral equations:
\begin{align}
\cL P_{\nu} (x;\ga)&=-\nu_{123}(\nu_{123}+\ga_{1234}+3)P_{\nu} (x;\ga),\label{4.4}\\
\cL_{2,3,4} P_{\nu} (x;\ga)&=-\nu_{23}(\nu_{23}+\ga_{234}+2)P_{\nu} (x;\ga),\label{4.5}\\
\cL_{3,4} P_{\nu} (x;\ga)&=-\nu_3(\nu_3+\ga_{34}+1)P_{\nu} (x;\ga).\label{4.6}
\end{align}
Let us define the space $\cV^{3}_n(\ga)$ spanned by the three-variable polynomials in \eqref{4.2} of total degree $n\in\Nset_0$, i.e. we set
\begin{equation}\label{4.7}
\cV^{3}_n(\ga)=\Span\{P_{\nu}(x;\ga):\nu_1+\nu_2+\nu_3=n\}.
\end{equation}
This space is a module of the symmetry algebra $\fA_3(\ga)$ generated by the operators $\cL_{i,j}$, $1\leq i<j\leq 4$. To describe the action of the remaining operators, similarly to the previous section, we use the shift operators $E_{\nu_1}$, $E_{\nu_2}$, $E_{\nu_3}$ acting on functions depending on $\nu=(\nu_1,\nu_2,\nu_3)$. First, we define the difference operator 
\begin{equation}\label{4.8}
\cB_{2,3}=b_{2,3}^{(0,-1,1)}(\nu)E_{\nu_2}^{-1}E_{\nu_3}+b_{2,3}^{(0,0,0)}(\nu)\Id+b_{2,3}^{(0,1,-1)}(\nu)E_{\nu_2}E_{\nu_3}^{-1},
\end{equation}
with coefficients
\begin{align*}
b_{2,3}^{(0,-1,1)}(\nu)&=\frac{\nu_2 (\ga_3+\nu_3+1) (\ga_{34}+\nu_3+1) (\ga_{234}+\nu_2+2 \nu_3+2)}{(\ga_{34}+2 \nu_3+1) (\ga_{34}+2 \nu_3+2)},\\
b_{2,3}^{(0,1,-1)}(\nu)&=\frac{\nu_3 (\ga_2+\nu_2+1) (\ga_4+\nu_3) (\ga_{34}+\nu_2+2 \nu_3+1)}{(\ga_{34}+2 \nu_3) (\ga_{34}+2 \nu_3+1)},\\
b_{2,3}^{(0,0,0)}(\nu)&=-(\nu_2+\nu_3+2\nu_2 \nu_3+\nu_2 \ga_3+\nu_3\ga_2 )\\
&\qquad +\frac{(\nu_2+1) \nu_3 (\ga_2+\nu_2+1) (\ga_3+\nu_3)}{\ga_{34}+2 \nu_3}
-\frac{\nu_2 (\nu_3+1) (\ga_2+\nu_2) (\ga_3+\nu_3+1)}{\ga_{34}+2\nu_3+2}.
\end{align*}
Then we have
\begin{equation} \label{4.9}
\cL_{2,3} P_{\nu}(x;\ga) =\cB_{2,3} P_{\nu}(x;\ga).
\end{equation} 
Similarly, if we define the difference operator
\begin{equation}\label{4.10}
\cB_{1,3,4}=b_{1,3,4}^{(1,-1,0)}(\nu)E_{\nu_1}E_{\nu_2}^{-1}+b_{1,3,4}^{(0,0,0)}(\nu)\Id+b_{1,3,4}^{(-1,1,0)}(\nu)E_{\nu_1}^{-1}E_{\nu_2},
\end{equation}
with coefficients
\begin{align*}
b_{1,3,4}^{(1,-1,0)}(\nu)&=-\frac{\nu_2 (\ga_1+\nu_1+1) (\ga_{34}+\nu_2+2 \nu_3+1) (\ga_{234}+\nu_1+2 \nu_{23}+2)}{(\ga_{234}+2 \nu_{23}+1) (\ga_{234}+2 \nu_{23}+2)},\\
b_{1,3,4}^{(-1,1,0)}(\nu)&=-\frac{\nu_1 (\ga_2+\nu_2+1) (\ga_{234}+\nu_2+2 \nu_3+2) (\ga_{1234}+\nu_1+2 \nu_{23}+3)}{(\ga_{234}+2 \nu_{23}+2) (\ga_{234}+2 \nu_{23}+3)},\\
b_{1,3,4}^{(0,0,0)}(\nu)&=-\nu_{13} (\ga_{134}+\nu_{13}+2)\\
&\qquad 
+\frac{\nu_1 (\nu_2+1) (\ga_1+\nu_1) (\ga_2+\nu_2+1)}{\ga_{234}+2 \nu_{23}+3}\
-\frac{(\nu_1+1) \nu_2 (\ga_1+\nu_1+1) (\ga_2+\nu_2)}{\ga_{234}+2 \nu_{23}+1},
\end{align*}
then 
\begin{equation} \label{4.11}
\cL_{1,3,4} P_{\nu}(x;\ga) =\cB_{1,3,4} P_{\nu}(x;\ga).
\end{equation} 
Finally, if we define the difference operator 
\begin{equation}\label{4.12}
\begin{split}
\cB_{1,2,3}&=b_{1,2,3}^{(-1,0,1)}(\nu)E_{\nu_1}^{-1}E_{\nu_3}+b_{1,2,3}^{(-1,2,-1)}(\nu)E_{\nu_1}^{-1}E_{\nu_2}^{2}E_{\nu_3}^{-1}+b_{1,2,3}^{(1,-2,1)}(\nu)E_{\nu_1}E_{\nu_2}^{-2}E_{\nu_3}\\ &+b_{1,2,3}^{(1,0,-1)}(\nu)E_{\nu_1}E_{\nu_3}^{-1} 
+b_{1,2,3}^{(0,-1,1)}(\nu)E_{\nu_2}^{-1}E_{\nu_3}+b_{1,2,3}^{(0,1,-1)}(\nu)E_{\nu_2}E_{\nu_3}^{-1}\\
&+b_{1,2,3}^{(-1,1,0)}(\nu)E_{\nu_1}^{-1}E_{\nu_2}+b_{1,2,3}^{(1,-1,0)}(\nu)E_{\nu_1}E_{\nu_2}^{-1}+b_{1,2,3}^{(0,0,0)}(\nu)\Id,
\end{split}
\end{equation}
with coefficients
\begin{align*}
b_{1,2,3}^{(-1,0,1)}(\nu)&=
\frac{\nu_1 (\ga_3+\nu_3+1) (\ga_{34}+\nu_3+1) (\ga_{234}+\nu_2+2 \nu_3+2) (\ga_{234}+\nu_2+2 \nu_3+3) }{(\ga_{34}+2 \nu_3+1) (\ga_{34}+2 \nu_3+2) (\ga_{234}+2 \nu_{23}+2) (\ga_{234}+2\nu_{23}+3)}\\
&\qquad\qquad\times (\ga_{1234}+\nu_1+2 \nu_{23}+3),\\ 
b_{1,2,3}^{(-1,2,-1)}(\nu)&=\frac{\nu_1\nu_3 (\ga_2+\nu_2+1) (\ga_2+\nu_2+2) (\ga_4+\nu_3) (\ga_{1234}+\nu_1+2 \nu_{23}+3)}{(\ga_{34}+2 \nu_3) (\ga_{34}+2 \nu_3+1) (\ga_{234}+2 \nu_{23}+2) (\ga_{234}+2\nu_{23}+3)},\\
b_{1,2,3}^{(1,-2,1)}(\nu)&=
\frac{(\nu_2-1) \nu_2 (\ga_1+\nu_1+1) (\ga_3+\nu_3+1) (\ga_{34}+\nu_3+1) (\ga_{234}+\nu_1+2 \nu_{23}+2)}{(\ga_{34}+2 \nu_3+1) (\ga_{34}+2 \nu_3+2) (\ga_{234}+2 \nu_{23}+1) (\ga_{234}+2 \nu_{23}+2)},\\
b_{1,2,3}^{(1,0,-1)}(\nu)&=\frac{\nu_3 (\ga_1+\nu_1+1) (\ga_4+\nu_3) (\ga_{34}+\nu_2+2 \nu_3) (\ga_{34}+\nu_2+2 \nu_3+1) }{(\ga_{34}+2 \nu_3) (\ga_{34}+2 \nu_3+1) (\ga_{234}+2 \nu_{23}+1) (\ga_{234}+2 \nu_{23}+2)}\\
&\qquad\qquad\times(\ga_{234}+\nu_1+2 \nu_{23}+2),\\
b_{1,2,3}^{(0,-1,1)}(\nu)&=
\nu_2 (\ga_{3}+\nu_3+1) (\ga_{34}+\nu_3+1) (\ga_{234}+\nu_2+2 \nu_3+2) \\
&\qquad\times 
\frac{\left(2\nu_{123} (\ga_{1234}+\nu_{123}+3)+2\nu_{23} (\ga_{234}+\nu_{23}+2)+(\ga_{234}+3) (\ga_{1234}+2)\right)}{(\ga_{34}+2 \nu_3+1) (\ga_{34}+2 \nu_3+2) \left((\ga_{234}+2 \nu_{23}+2)^2-1\right)},\\
b_{1,2,3}^{(0,1,-1)}(\nu)&=\nu_3 (\ga_{2}+\nu_2+1) (\ga_{4}+\nu_3) (\ga_{34}+\nu_2+2 \nu_3+1)\\
&\qquad\times \frac{ \left(2 \nu_1 (\ga_{1234}+\nu_1+2\nu_{23}+3)+(\ga_{234}+2\nu_{23}+3) (\ga_{1234}+2\nu_{23}+2)\right)}{(\ga_{34}+2 \nu_3) (\ga_{34}+2 \nu_3+1) (\ga_{234}+2\nu_{23}+1) (\ga_{234}+2\nu_{23}+3)},\\
%\end{align*}
%\begin{align*}
b_{1,2,3}^{(-1,1,0)}(\nu)&=\frac{\nu_1 (\ga_{2}+\nu_2+1)  (\ga_{234}+\nu_2+2 \nu_3+2) (\ga_{1234}+\nu_1+2\nu_{23}+3)}{\left((\ga_{34}+2 \nu_3+1)^2-1\right) (\ga_{234}+2 \nu_{23}+2) (\ga_{234}+2 \nu_{23}+3)}\\
&\qquad\qquad\times \left(2\nu_3 (\ga_{34}+\nu_3+1)+(\ga_{4}+1) \ga_{34}\right),\\
b_{1,2,3}^{(1,-1,0)}(\nu)&=\frac{\nu_2 (\ga_1+\nu_1+1) (\ga_{34}+\nu_2+2 \nu_3+1) (\ga_{234}+\nu_1+2\nu_{23}+2)}{(\ga_{34}+2 \nu_3) (\ga_{34}+2 \nu_3+2) (\ga_{234}+2\nu_{23}+1) (\ga_{234}+2\nu_{23}+2)}\\
&\qquad\qquad\times \left(2\nu_3 (\ga_{34}+\nu_3+1)+(\ga_{4}+1) \ga_{34}\right),\\
%\end{align*}
%\begin{align*}
b_{1,2,3}^{(0,0,0)}(\nu)&=-\nu_{123} (\ga_{1234}+\nu_{123}+3)-\frac{1}{2} (\ga_{4}+1) (\ga_{1234}+2)\\
&+\frac{(2\nu_3 (\ga_{34}+\nu_3+1)+(\ga_{4}+1) \ga_{34}) }
{2 (\ga_{34}+2 \nu_{3}) (\ga_{34}+2 \nu_{3}+2)}\\
&\qquad\times \left(2 \nu_{2} (\ga_{234}+\nu_2+2 \nu_{3}+2)+(\ga_{34}+2 \nu_{3}+2) (\ga_{234}+2 \nu_{3}+1)\right)\\
&\qquad\times\frac{\left(2 \nu_{1} (\ga_{1234}+\nu_1+2\nu_{23}+3)+(\ga_{234}+2\nu_{23}+3) (\ga_{1234}+2\nu_{23}+2)\right)}{ (\ga_{234}+2\nu_{23}+1) (\ga_{234}+2\nu_{23}+3)},
\end{align*}
then 
\begin{equation} \label{4.13}
\cL_{1,2,3} P_{\nu}(x;\ga) =\cB_{1,2,3} P_{\nu}(x;\ga).
\end{equation} 
Equations \eqref{4.4}, \eqref{4.5}, \eqref{4.6}, \eqref{4.9}, \eqref{4.11} and \eqref{4.13} describe the action of the operators $\cL$, $\cL_{2,3,4}$, $\cL_{3,4}$, $\cL_{2,3}$, $\cL_{1,3,4}$, $\cL_{1,2,3}$ on the basis $\{P_{\nu}(x;\ga)\}_{\nu_{123}=n}$ of $\cV^{3}_n(\ga)$. The action of all other second-order generators of the symmetry algebra can be easily computed from the relations
\begin{equation}\label{4.14}
\begin{split}
\cL_{1,2}&=\cL-\cL_{1,3,4}-\cL_{2,3,4}+\cL_{3,4},\\
\cL_{1,3}&=\cL_{1,2,3}+\cL_{1,3,4}+\cL_{2,3,4}-\cL-\cL_{2,3}-\cL_{3,4},\\
\cL_{1,4}&=\cL-\cL_{1,2,3}-\cL_{2,3,4}+\cL_{2,3},\\
\cL_{2,4}&=\cL_{2,3,4}-\cL_{2,3}-\cL_{3,4}.
\end{split}
\end{equation}
The representations of the symmetry algebra for the $3$-sphere first appeared in the work of Kalnins, Miller and Post \cite{KMP2}, where the connection with the two-variable Racah operators constructed in \cite{GI} for the bivariate Racah polynomials introduced by Tratnik \cite{Tr} was discovered. The connection with the Jacobi polynomials~\eqref{4.2} outlined here stems from the approach in \cite{I} and can be explained as follows.  If $\ga_j>-1$ for $j=1,2,3,4$, then the polynomials $P_{\nu}(x;\ga)$ are mutually orthogonal with respect to the weight 
$$w_3(x_1,x_2,x_3)=x_1^{\ga_1}x_2^{\ga_2}x_3^{\ga_3}(1-x_1-x_2-x_3)^{\ga_4}$$
on the simplex $\Tset^3=\{(x_1,x_2,x_3)\in\Rset^3:0\leq x_i,\, x_1+x_2+x_3\leq 1\}$. The symmetry operators $\cL_{1,2}$, $\cL_{1,3}$, $\cL_{1,4}$, $\cL_{2,3}$, $\cL_{2,4}$, $\cL_{3,4}$ are self-adjoint with respect to the inner product induced by $w_3$. This implies that the symmetry algebra will preserve the space $\cV^{3}_n(\ga)$ of orthogonal polynomials of degree total $n$. Note also that $w_3$ is invariant if we permute simultaneously $(\ga_1,\ga_2,\ga_3,\ga_4)$ and $(x_1,x_2,x_3,x_4)$, where $x_4=1-x_1-x_2-x_3$. This shows that we can construct other bases of $\cV^{3}_n(\ga)$ by applying permutations.  The key point now is that the transition matrices between these different orthogonal bases can be expressed in terms of the Racah weight and polynomials of one and two variables, see \cite[Section 5]{IX}.

\subsection{Irreducibility}\label{ss4.2}
The goal of this subsection is to show that $\cV^{3}_n(\ga)$ is an irreducible module over $\fA_3(\ga)$.
\subsubsection{Preliminary observations}
For $\nu\in\Nset_0^3$ we set 
$$\la_1(\nu)=-\nu_{123}(\nu_{123}+\ga_{1234}+3), \quad \la_2(\nu)=-\nu_{23}(\nu_{23}+\ga_{234}+2), \quad \la_3(\nu)=-\nu_{3}(\nu_{3}+\ga_{34}+1).$$
Computations similar to the ones for the $2$-sphere show that 
\begin{equation}\label{4.15}
(\la_1(\nu),\la_2(\nu),\la_3(\nu))=(\la_1(\mu),\la_2(\mu),\la_3(\mu))\text{ if and only if }\nu=\mu.
\end{equation}
Therefore, the spectral equations \eqref{4.4}, \eqref{4.5}, \eqref{4.6} fix the polynomials  $P_{\nu}(x;\ga)$, up to unessential factors depending only on $\nu$.
Next, we note that the module $\cV^{3}_n(\ga)$ contains several copies of the modules described for the $2$-sphere. 

First, we consider the operators $\cL_{2,3}$, $\cL_{2,4}$, $\cL_{3,4}$ which generate the algebra $\fA_2(\ga_2,\ga_3,\ga_4)$. Since these operators do not contain differentiation with respect to $x_1$, it follows that they commute with the first factor on the right-hand side of equation~\eqref{4.2}. Therefore, if we fix $k,n\in\Nset_0$ such that $0\leq k\leq n$, then the operators $\cL_{2,3}$, $\cL_{2,4}$, $\cL_{3,4}$ preserve the subspace 
\begin{equation}\label{4.16}
\cW^{1}_{n,k}(\ga)=\Span\{P_{k,\nu_2,\nu_3}(x;\ga):\nu_2+\nu_3=n-k\},
\end{equation}
of $\cV^{3}_{n}(\ga)$. It is not hard to see now that $\cW^{1}_{n,k}(\ga)$ can be identified with the module $\cV^{2}_{n-k}(\ga_2,\ga_3,\ga_4)$ constructed in the previous section. One way to see this is to note that, up to a factor independent of $x_2$ and $x_3$, the product consisting of the last two terms in \eqref{4.2}:
$$p_{\nu_2}^{(\ga_{34}+2\nu_3+1,\ga_2)}\left (\frac{2 x_2}{1-x_1}-1\right)
(1-x_1-x_2)^{\nu_3} p_{\nu_3}^{(\ga_4,\ga_3)}\left (\frac{2 x_3}{1-x_1-x_2}-1\right)$$
coincides with the two-variable polynomial $P_{\nu_2,\nu_3}(y_1,y_2;\ga_2,\ga_3,\ga_4)$ in \eqref{3.2} in the variables $y_1$ and $y_2$ defined by
$$y_1=\frac{x_2}{1-x_1} \qquad\text{ and }\qquad  y_2=\frac{x_3}{1-x_1}.$$
Moreover, if make the same change of variables in the operators $\cL_{2,3}$, $\cL_{2,4}$, $\cL_{3,4}$ for the 3-sphere, we obtain in the $y$ variables the operators  $\cL_{1,2}$, $\cL_{1,3}$, $\cL_{2,3}$ for the 2-sphere. It is also useful to compare equations  \eqref{4.9}, \eqref{4.6}, \eqref{4.5} in the three-dimensional setting with equations \eqref{3.7}, \eqref{3.5}, \eqref{3.4} in the two-dimensional setting, respectively. 

Next, we consider the subspace 
\begin{equation}\label{4.17}
\cW^{0}_{n}(\ga)=\Span\{P_{\nu_1,\nu_2,0}(x;\ga):\nu_1+\nu_2=n\},
\end{equation}
of $\cV^{3}_{n}(\ga)$. We want to show that the operators $\cLh_{1,2}=\cL_{1,2}$, $\cLh_{1,3}=\cL_{1,3}+\cL_{1,4}$,  $\cLh_{2,3}=\cL_{2,3}+\cL_{2,4}$ preserve this space, and we can identify the subspace $\cW^{0}_{n}(\ga)$ under their action with the module $\cV^{2}_{n}(\ga_1,\ga_2,\ga_3+\ga_4+1)$ constructed in the previous section. Clearly, $\cW^{0}_{n}(\ga)$ is a space of polynomials in the variables $x_1$ and $x_2$ and therefore $\pd_{x_3}$ will act as the zero operator on this space. Using equation \eqref{2.5} with $i=1$, $j=3$ and ignoring $\pd_{x_3}$, we see that 
\begin{equation*}
\cL_{1,3}\Big|_{\cW^{0}_{n}(\ga)}=x_1x_3\pd_{x_1}^2+[(\ga_1+1)x_3-(\ga_3+1)x_1]\pd_{x_1}.
\end{equation*}
Similarly, using equation \eqref{2.6} with $j=1$ and $d=3$ we deduce
\begin{equation*}
\cL_{1,4}\Big|_{\cW^{0}_{n}(\ga)}=x_1(1-x_1-x_2-x_3)\pd_{x_1}^2+[(\ga_1+1)(1-x_1-x_2-x_3)-(\ga_{4}+1)x_1]\pd_{x_1}.
\end{equation*}
Adding the last two equations, we see that 
$$\cLh_{1,3}\Big|_{\cW^{0}_{n}(\ga)}=x_1(1-x_1-x_2)\pd_{x_1}^2+[(\ga_1+1)(1-x_1-x_2)-(\ga_3+\ga_{4}+2)x_1]\pd_{x_1},$$
which coincides with the operator in $\cL_{1,3}$ defined by \eqref{2.6} in dimension $d=2$ with parameter $\ga_3$ replaced by $\ga_3+\ga_4+1$.
A similar computation shows that $\cLh_{2,3}\Big|_{\cW^{0}_{n}(\ga)}$ coincides with the operator in $\cL_{2,3}$ defined by \eqref{2.6} in dimension $d=2$ with parameter $\ga_3$ replaced by $\ga_3+\ga_4+1$. Finally, it is easy to see that if we put $\nu_3=0$ in equation \eqref{4.2}, the expression on the second line is 1 and $P_{\nu_1,\nu_2,0}(x;\ga)$ coincides with the two-variable polynomials defined by \eqref{3.2} with  parameter $\ga_3$ replaced by $\ga_3+\ga_4+1$. Applying the results in Section~\ref{se3}, we conclude that $\cW^{0}_{n}(\ga)$ is an irreducible module over the algebra generated by  $\cLh_{1,2}$, $\cLh_{1,3}$,  $\cLh_{2,3}$.

\subsubsection{Proof of the irreducibility} 
Using the observations above we prove below that $\cV^{3}_n(\ga)$ is an irreducible module over $\fA_3(\ga)$. We fix $n\in\Nset$ and suppose that $\cV$ is a nontrivial submodule of $\cV^{3}_n(\ga)$.  Equations \eqref{4.5}, \eqref{4.6} and \eqref{4.15} show that the operators $\cL_{2,3,4}$ and $\cL_{3,4}$ can be simultaneously diagonalized on $\cV$. Therefore, there exists at least one polynomial $P_{k_1,k_2,k_3}(x;\ga)$ with $k_1+k_2+k_3=n$ which belongs to $\cV$. Since $P_{k_1,k_2,k_3}(x;\ga)\in \cW^{1}_{n,k_1}(\ga)$, and using the irreducibility of $ \cW^{1}_{n,k_1}(\ga)$ established above, we conclude that $\cW^{1}_{n,k_1}(\ga)\subset\cV$. In particular, this means that 
$P_{k_1,n-k_1,0}(x;\ga)\in\cV$. On the other hand, $P_{k_1,n-k_1,0}(x;\ga)\in\cW^{0}_{n}(\ga)$. Using now the irreducibility of $\cW^{0}_{n}(\ga)$, as a module over the algebra generated by  $\cLh_{1,2}$, $\cLh_{1,3}$,  $\cLh_{2,3}$, we deduce that $\cW^{0}_{n}(\ga)\subset\cV$. Therefore, for every $\nu_1\in\Nset_0$ such that $0\leq \nu_1\leq n$, we have $P_{\nu_1,n-\nu_1,0}(x;\ga)\in \cV$. Finally, we use the fact that $P_{\nu_1,n-\nu_1,0}(x;\ga)\in   \cW^{1}_{n,\nu_1}(\ga)$ and therefore  $\cW^{1}_{n,\nu_1}(\ga)\subset\cV$. Since this is true for every $\nu_1\in\{0,1,\dots,n\}$ we conclude that $\cV=\cV^{3}_n(\ga)$, completing the proof.

\section{Symmetry algebra for the $d$-sphere}\label{se5}
Recall that $\fA_d(\ga)$ is the algebra generated by the operators $\cL_{i,j}$, where $1\leq i<j\leq d+1$. For every $n\in\Nset_0$, we construct in this section a module over $\fA_d(\ga)$ consisting of polynomials of total degree $n$ in $d$ variables, and we show that this module is irreducible.

\subsection{Construction of the module} \label{ss5.1}

\subsubsection{Jacobi polynomials in $d$ variables and a commuting subfamily of operators}
We start by introducing some notations which will help us write explicit formulas in arbitrary dimension. For a vector $v=(v_1,\dots, v_{s})$
%we denote by 
%$$|v|=v_1+\cdots+v_s$$ 
%the sum of its components, 
and for $j=0,1,\dots,s+1$ we define 
\begin{equation*}
    \bv_j = (v_1, \dots, v_j) \quad \text{ and }\quad \bv^j = (v_j, \ldots, v_s), 
\end{equation*}
with the convention that $\bv_0 =\bv^{s+1} = 0$.
Following \cite{DX}, we define polynomials in $x=(x_1,\dots,x_d)$, with degree indices $\nu=(\nu_1,\dots,
\nu_d)$, and depending on the parameters $\ga=(\ga_1,\dots,\ga_{d+1})$, in terms of the one-variable Jacobi polynomials as
\begin{equation}\label{5.1}
P_{\nu}(x;\ga) = \prod_{k=1}^d \left(1-|\bx_{k-1}| \right)^{\nu_k} 
               p_{\nu_k}^{(a_k,\ga_k)}\left (\frac{2x_k}{1-|\bx_{k-1}|} -1\right), 
\end{equation}
where
\begin{equation}\label{5.2}
a_j=a_j(\ga,\nu)=|\bga^{j+1}| + 2 |\bnu^{j+1}| + d-j, \qquad 1 \le j \le d.
\end{equation} 
Note that these polynomials are well-defined if the parameters satisfy the following conditions
\begin{equation}\label{5.3}
\begin{split}
\ga_j\notin \Zset_{\leq -1}\quad \text{ and } \quad|\bga^{j}|\notin \Zset_{\leq -d+j-2}\quad \text{ for }j=1,\dots,d+1.
\end{split}
\end{equation}
It is easy to see that the polynomials in \eqref{5.1} reduce to the two-variable polynomials in \eqref{3.2} when $d=2$, and to the three-variable polynomials in \eqref{4.2} when $d=3$. Similarly, the conditions \eqref{5.3} on the parameters extends equations \eqref{3.3} and \eqref{4.3}. In order to generalize the spectral equations, it is convenient to introduce also the operators
\begin{equation}\label{5.4}
\cM_{j}=\sum_{j\leq k<l\leq d+1}\cL_{k,l}, \qquad \text{ for }\qquad j=1,2,\dots,d.
\end{equation}
From \eqref{2.9} and \eqref{2.10}, it follows that the operators $\cM_{j}$ commute with each other, i.e. 
$$[\cM_{i},\cM_{j}]=0.$$
Moreover, $\cL=\cM_{1}$.
The polynomials $P_{\nu}(x;\ga) $ satisfy the spectral equations
\begin{equation}\label{5.5}
\cM_{j}P_\nu (x;\ga)=-|\bnu^{j}|(|\bnu^{j}|+|\bga^{j}|+d+1-j)P_\nu (x;\ga), \text{ for }j=1,\dots,d.
\end{equation}
For $n\in\Nset_0$,  we define the space $\cV^{d}_n(\ga)$ spanned by the polynomials in \eqref{5.1} of total degree $n$, i.e. we set
\begin{equation}\label{5.6}
\cV^{d}_n(\ga)=\Span\{P_{\nu}(x;\ga):\nu_1+\cdots+\nu_d=n\}.
\end{equation}
We will show that $\cV^{d}_n(\ga)$ is an irreducible module of the algebra  $\fA_{d}(\ga)$ generated by the symmetry operators $\cL_{i,j}$, $1\leq i<j\leq d+1$. We postpone the proof of the irreducibility for later, and we describe first the action of all operators $\cL_{i,j}$ on the basis $\{P_{\nu}(x;\ga)\}$.

\subsubsection{Racah operators}
Consider variables $z=(z_1,z_2,\dots)$ and parameters $\be=(\be_0,\be_1,\dots)$.  We extend the convention in the previous sections  and for every $j\in\Nset$ we set $\bz_j=(z_1,\dots,z_j)$ and $\bbe_j=(\be_0,\be_1,\dots,\be_j)$. Moreover, we can consider every finite-dimensional vector as a semi-infinite vector with zeros after the last component.

We denote by $\Rset(z)$ the field of rational functions of finitely many of the $z_j$'s and for $k\in\Nset$ we define an involution $I_k$ on $\Rset(z)$, by 
\begin{equation*}
I_k(z_k)=-z_k-\be_k \text{ and }I_k(z_j)=z_j \text{ for }j\neq k.
\end{equation*}
For $k\in\Nset$ we denote by $E_{z_k}$ the forward shift operator acting on the variable $z_k$, i.e. if $f(z)\in\Rset(z)$ then 
\begin{equation*}
E_{z_k}f(z_1,z_2,\dots,z_{k-1},z_{k},z_{k+1},\dots)=f(z_1,z_2,\dots,z_{k-1},z_{k}+1,z_{k+1},\dots).
\end{equation*}
The inverse  $E_{z_k}^{-1}$ of  $E_{z_k}$ corresponds to the backward shift in the variable $z_k$ defined by
\begin{equation*}
E_{z_k}^{-1}f(z_1,z_2,\dots,z_{k-1},z_{k},z_{k+1},\dots)=f(z_1,z_2,\dots,z_{k-1},z_{k}-1,z_{k+1},\dots).
\end{equation*}
Let $\Zset^{\infty}=\{(\nu_1,\nu_2,\dots): \nu_j\neq 0 \text{ for finitely many }j\}$ be the additive group consisting of semi-infinite vectors having finitely many nonzero integer entries. For $\nu\in\Zset^{\infty}$ we have a well-defined shift operator
$$E_z^{\nu}=E_{z_1}^{\nu_1}E_{z_2}^{\nu_2}E_{z_3}^{\nu_3}\cdots,$$
since the right-hand side has only finitely many terms different from the identity operator.
We denote by $\cDz$ the associative algebra of difference operators of the form 
$$L=\sum_{\nu\in S}l_\nu(z)E_z^{\nu},$$
where $S$ is a finite subset of $\Zset^{\infty}$ and $l_{\nu}(z)\in\Rset(z)$. The involution $I_k$ can be extended to an involution on $\cDz$ by defining 
\begin{equation*}
I_k(E_{z_k})=E_{z_k}^{-1} \text{ and }I_k(E_{z_j})=E_{z_j} \text{ for }j\neq k.
\end{equation*}
We say that an operator $L\in\cDz$ is $I$-invariant, if it is invariant under the action of all involutions $I_k$, $k\in\Nset$. 

Following \cite{GI} we define below a commutative subalgebra of $\cDz$ consisting of $I$-invariant Racah operators. 
For $i\in\Nset_0$ and $(j,k)\in\{0,1\}^2$ we define $B_{i}^{j,k}$
as follows
\begin{align*}
B_i^{0,0}&=z_{i}(z_{i}+\be_{i})+z_{i+1}(z_{i+1}+\be_{i+1})+\frac{(\be_{i}+1)(\be_{i+1}-1)}{2},\\
B_i^{0,1}&=(z_{i+1}+z_{i}+\be_{i+1})(z_{i+1}-z_{i}+\be_{i+1}-\be_{i}),\\
B_i^{1,0}&=(z_{i+1}-z_{i})(z_{i+1}+z_{i}+\be_{i+1}),\\
B_i^{1,1}&=(z_{i+1}+z_{i}+\be_{i+1})(z_{i+1}+z_{i}+\be_{i+1}+1),
\end{align*}
where $z_0=0$.
For $i\in\Nset$ we denote
\begin{align*}
b_i^{0}&=\frac{(2z_{i}+\be_i+1)(2z_{i}+\be_i-1)}{2},\\
b_i^{1}&=(2z_{i}+\be_i+1)(2z_{i}+\be_i).
\end{align*}
Using the above notations, for $j\in\Nset$ and $\nu\in\{0,1\}^j$ we define 
\begin{equation*}
C_{j,\nu}(z)=\frac{\prod_{k=0}^{j}B_k^{{\nu}_k,{\nu}_{k+1}}}
{\prod_{k=1}^{j}b_k^{{\nu}_k}}, 
\end{equation*}
where $\nu_0=\nu_{j+1}=0$. We extend the definition of $C_{j,\nu}$ for $\nu\in\{-1,0,1\}^j$ using the involutions $I_k$ as follows. 
Every $\nu\in\{-1,0,1\}^j$ can be decomposed as $\nu=\nu^{+}-\nu^{-}$, 
where $\nu^{\pm}\in \{0,1\}^j$ with components $\nu_k^{+}=\max(\nu_k,0)$ and 
$\nu_k^{-}=-\min(\nu_k,0)$.
For $\nu\in\{-1,0,1\}^j\setminus \{0,1\}^j$ we 
define 
\begin{equation*}
C_{j,\nu}(z)=I^{\nu^{-}}(C_{j,\nu^{+}+\nu^{-}}(z)),
\end{equation*}
where $I^{\nu^{-}}$ is the composition of the involutions corresponding to the positive coordinates of $\nu^{-}$.
Finally, for $j\in\Nset$ we define 
\begin{equation}\label{5.7}
\cB_j(z;\be)=\sum_{\nu\in\{-1,0,1\}^j}C_{j,\nu}(z)E_z^{\nu}
-\left(z_{j+1}(z_{j+1}+\be_{j+1})+\frac{(\be_{0}+1)(\be_{j+1}-1)}{2}\right)\Id.
\end{equation}
Note that $\cB_j(z;\be)$ are $I$-invariant difference operators in the variables $\bz_{j}=(z_1,\dots,z_{j})$ 
with coefficients depending rationally on $\bz_{j+1}=(z_1,\dots,z_{j+1})$ and $\bbe_{j+1}=(\be_0,\be_1,\dots,\be_{j+1})$.
The operators $\cB_j(z;\be)$ commute with each other, i.e.
$$[\cB_j(z;\be),\cB_k(z;\be)]=0,$$
see \cite[Section 3]{GI}.

\subsubsection{Action of generators for $\fA_d(\ga)$} 

For any permutation $\tau$ of the set $\{1,2,\dots,d+1\}$ we define $\tau(\cL_{i,j})=\cL_{\tau(i),\tau(j)}$. We fix below $\tau$ to be the cyclic permutation
$$\tau=(1,2,\dots,d+1),$$
and we set $\cM_j^{+}$ and $\cM_j^{-}$ to be the operators obtained from the operator $\cM_j$ in \eqref{5.4} by applying $\tau$ and $\tau^{-1}$, respectively, i.e.
$$\cM_j^{+}=\tau(\cM_j)=\sum_{j\leq k<l\leq d+1}\cL_{\tau(k),\tau(l)},$$
and 
$$\cM_j^{-}=\tau^{-1}(\cM_j)=\sum_{j\leq k<l\leq d+1}\cL_{\tau^{-1}(k),\tau^{-1}(l)},$$
for $j=1,\dots,d$. Clearly, $\cM_1^{+}=\cM_1^{-}=\cM_1$. The action of the operators $\cM_{j}^{\pm}$ for $j=2,\dots,d$ can be expressed in terms of the Racah operators defined in \eqref{5.7}.
More precisely, for fixed $n\in\Nset_0$ we define two sets of parameters
\begin{equation}\label{5.8}
\be^{+}_0=\ga_1, \quad \be^{+}_j=-|\bga^{j+1}|-2n-d+j,\qquad\text{ for }j=1,\dots,d,
\end{equation}
and 
\begin{equation}\label{5.9}
\be^{-}_j=|\bga^{d+1-j}|+j,\qquad\text{ for }j=0,1,\dots,d.
\end{equation}
For $\nu=(\nu_1,\dots,\nu_d)$ we set 
$$\nu_{+}=(|\bnu_1|,|\bnu_{2}|,\dots,|\bnu_{d-1}|,|\bnu_{d}|)\qquad\text{ and }\qquad \nu_{-}=(|\bnu^d|,|\bnu^{d-1}|,\dots,|\bnu^{2}|,|\bnu^1|).$$
Then for every $\nu\in\Nset_0^d$ with $|\nu|=n$ and $j=2,\dots, d$ we have
\begin{align}
&\cM^{+}_{j}P_{\nu}(x;\ga)%\nonumber\\
%&\quad
=\left(n(n+\be^{+}_{j}-\be^{+}_0-1)+\frac{1}{g(\nu;\ga)}\cB_{j-1}(\nu_{+};\be^{+}) g(\nu;\ga)\right)P_{\nu}(x;\ga),\label{5.10}
\intertext{and}
&\cM^{-}_{j} P_{\nu}(x;\ga)=\cB_{d+1-j}(\nu_{-};\be^{-})P_{\nu}(x;\ga),\label{5.11}
\end{align}
where  
\begin{equation}\label{5.12}
g(\nu;\ga)=\frac{(1+\ga_1)_{\nu_1}}{(|\ga|+2n+d-\nu_1)_{\nu_1}}.
\end{equation}
The operator $\cB_{j-1}(\nu_{+};\be^{+})$ in \eqref{5.10} is a difference operator in the variables $\nu_{1},\dots,\nu_{d}$ obtained from the operator in \eqref{5.7} by changing the variables. Explicitly, we replace $z_l$ by $|\bnu_{l}|=\nu_1+\cdots+\nu_l$ in the coefficients, and we replace $E_{z_l}$ by $E_{\nu_{l}}E_{\nu_{l+1}}^{-1}$. The operator $\cB_{d+1-j}(\nu_{-};\be^{-})$ in equation \eqref{5.11} is defined in a similar manner. Equations \eqref{5.10} and \eqref{5.11} follow from Proposition 5.1 in \cite{I}.
Note that these formulas extend the operators constructed in the previous sections. For instance, when $d=3$, equation \eqref{5.10} with $j=2$ corresponds to \eqref{4.11}, while equation \eqref{5.11} with $j=2$ and $j=3$ corresponds to equations \eqref{4.13} and \eqref{4.9}, respectively. 

\subsubsection{Action of all elements in $\fA_{d}(\ga)$} So far, we have described in equations \eqref{5.5}, \eqref{5.10} and \eqref{5.11} the action of the elements $\cM_j$, $\cM^{+}_j$ and $\cM^{-}_j$ on the basis $P_{\nu}(x;\ga)$. In dimensions $2$ and $3$, we can easily obtain the action of all elements in the symmetry algebra, by taking appropriate linear combinations of the elements $\cM_j$, $\cM^{\pm}_j$. Indeed, if $d=3$, then 
$$\cM_1=\cL,\; \cM_2=\cL_{2,3,4},\; \cM_{3}=\cL_{3,4},\; \cM^{+}_2=\cL_{1,3,4},\; \cM^{-}_2=\cL_{1,2,3},\; \cM^{-}_{3}=\cL_{2,3},$$ 
and we can compute the action of all other elements $\cL_{i,j}$ by using \eqref{4.14}. In dimension $d\geq 4$ it is no longer possible to use only linear combinations to compute $\cL_{i,j}$ in terms of the operators $\cM_j$, $\cM^{\pm}_{j}$. To see why this is true, note first that the operators  $\cL_{i,j}$ are linearly independent  (as operators on the space of polynomials) and their number is $\binom{d+1}{2}$. Indeed, to check the independence,  assume that there exist scalars $c_{i,j}$ such that
\begin{equation}\label{5.13}
\sum_{1\leq i<j\leq d+1}c_{i,j}\cL_{i,j}=0.
\end{equation}
Let us fix $i\in\{1,\dots,d\}$  and let us apply both sides of \eqref{5.13} to the monomial $x_i^{n_i}$. Computing first the coefficient of $x_i^{n_i-1}$ yields
$$c_{i,d+1}n_i(n_i+\ga_i)=0.$$
Since the last equation is true for all $n_i$, we see that $c_{i,d+1}=0$. Assuming now that $c_{i,d+1}=0$ for all $i\in\{1,\dots,d\}$, applying both sides of \eqref{5.13} to the monomial $x_i^{n_i}$ and computing the coefficient of $x_i^{n_i-1}x_j$ gives
$$c_{i,j}n_i(n_i+\ga_j)=0,$$
which shows that $c_{i,j}=0$ for every $j\neq i\in \{1,\dots,d\}$, completing the proof of the independence of $\cL_{i,j}$. 
Note next that the number of the operators in the set 
$$\{\cM_j:j=1,\dots,d\}\cup\{\cM_j^{+}:j=2,\dots,d\}\cup\{\cM_j^{-}:j=2,\dots,d\}$$ 
is $3d-2$. It is easy to see that the operators $\cM_j$, $\cM^{\pm}_{j}$ are linearly dependent, since
$$\cM_1-\cM_2=\cM^{-}_2-\cM^{-}_3+\cM^{+}_d.$$
Therefore, they span a space of dimension at most $3d-3$. Since the operators $\cL_{i,j}$ are linearly independent and $\binom{d+1}{2}>3d-3$ for $d\geq 4$, we see that linear combinations of the operators $\cM_j$, $\cM^{\pm}_j$ are not sufficient to derive explicit formulas for all $\cL_{i,j}$.

It turns out, however, that there is an explicit fourth-order algebraic relation which allows to express all $\cL_{i,j}$'s in terms of the operators $\cM_k$, $\cM^{\pm}_k$, $k=1,\dots,d$. First, note that we can obtain the operators $\cL_{1,j}$ and $\cL_{i,d+1}$ via the formulas
\begin{align}
\cL_{1,j}&=\cM^{+}_{j-1}+\cM_{j+1}-\cM_{j}-\cM^{+}_{j}, &&\text{ for }j=2,\dots,d+1,\label{5.14}\\
\cL_{i,d+1}&=\cM_{i}+\cM^{-}_{i+2}-\cM^{-}_{i+1}-\cM_{i+1}, &&\text{ for }i=1,\dots,d,\label{5.15}
\end{align}
with the convention that $\cM_{k}=\cM^{\pm}_{k}=0$ for $k=d+1,d+2$. Next, one can check that if $i,j,k,l$ are distinct, then 
\begin{equation}\label{5.16}
(1-\ga_k^2)(1-\ga_l^2)\cL_{i,j}=F(\cL_{i,k},\cL_{i,l},\cL_{j,k},\cL_{j,l},\cL_{k,l}),
\end{equation}
where
\begin{align}
&F(\cL_{i,k},\cL_{i,l},\cL_{j,k},\cL_{j,l},\cL_{k,l})\nonumber\\
&\qquad=\left \{ [\cL_{j,k},\cL_{k,l}],[\cL_{i,k},\cL_{k,l}] \right\}-\left\{\cL_{k,l},[\cL_{i,k},[\cL_{j,k},\cL_{k,l}]] \right\}\nonumber\\
&\qquad -2\left\{\cL_{k,l},\cL_{i,k}\cL_{j,l}\right\}
+(1+\ga_k)(1+\ga_l)\left[\cL_{i,k},[\cL_{k,l},\cL_{j,l}]\right]\nonumber\\
&\qquad+(1+\ga_j)(1+\ga_l)\left\{\cL_{i,k},\cL_{k,l}\right\}+(1-\ga_l^2)\left\{\cL_{i,k},\cL_{j,k}\right\}\nonumber\\
&\qquad+(1-\ga_k^2)\left\{\cL_{i,l},\cL_{j,l}\right\}+(1+\ga_i)(1+\ga_k) \left\{\cL_{j,l},\cL_{k,l}\right\}\nonumber\\
&\qquad-4\cL_{j,k}\cL_{i,l}+2(-1+\ga_k+\ga_l+\ga_k\ga_l)\cL_{j,l}\cL_{i,k}\nonumber\\
&\qquad-2\ga_k(1+\ga_j)(1+\ga_l) \cL_{i,k}+(1+\ga_j)(1+\ga_k)(1+\ga_k-\ga_l+\ga_k\ga_l)\cL_{i,l}\nonumber\\
&\qquad+(1+\ga_i)(1+\ga_l)(1-\ga_k+\ga_l+\ga_k\ga_l)\cL_{j,k}-2(1+\ga_i)(1+\ga_k)\ga_l\cL_{j,l}\nonumber\\
&\qquad -(1+\ga_i)(1+\ga_j)(1+\ga_k)(1+\ga_l)\cL_{k,l}.\label{5.17}
\end{align}
In the above formula, $\{A,B\}=AB+BA$ denotes the anticommutator of the operators $A$ and $B$.

Equation \eqref{5.16} provides and interesting link to the $5\Rightarrow 6$ Theorem in the general structure theory for 3D second-order superintegrable systems with nondegenerate potentials \cite{KKM}. According to this theorem, $5$ algebraically independent second-order symmetries guarantee the existence of an additional second-order symmetry, such that the $6$ symmetries are linearly independent and generate a quadratic algebra. If we think of the $6$ operators $\cL_{i,j},\cL_{i,k},\cL_{i,l},\cL_{j,k},\cL_{j,l},\cL_{k,l}$ as symmetries for the generic superintegrable system on the $3$-sphere, then \eqref{5.16} gives an explicit formula for each one of these symmetries in terms of the other five.

Note now that equations \eqref{5.5}, \eqref{5.10}, \eqref{5.11}, \eqref{5.14}, \eqref{5.15} and \eqref{5.16} allow us  to compute the action of all operators $\cL_{i,j}$ on the basis $P_{\nu}(x;\ga)$ of $\cV^d_n(\ga)$. First, we compute the action of the operators $\cL_{1,j}$ and $\cL_{i,d+1}$ using \eqref{5.5}, \eqref{5.10}, \eqref{5.11}, \eqref{5.14} and \eqref{5.15}. Then, for all $1<i<j<d+1$, we can compute the action of $\cL_{i,j}$ using \eqref{5.16} with $k=1$ and $l=d+1$, 
via the formula
$$\cL_{i,j}=\frac{1}{(1-\ga_1^2)(1-\ga_{d+1}^2)}F(\cL_{1,i},\cL_{i,d+1},\cL_{1,j},\cL_{j,d+1},\cL_{1,d+1}).$$
Since we assume that $\ga_k\neq -1$ for all $k$, the only subtle point when we use the above formula arises when $\ga_1=1$ or $\ga_{d+1}=1$. But there is a simple argument which shows that, after we substitute the Racah operators in $F(\cL_{1,i},\cL_{i,d+1},\cL_{1,j},\cL_{j,d+1},\cL_{1,d+1})$, each coefficient of the resulting difference operator must be divisible by $(1-\ga_1)(1-\ga_{d+1})$. Thus, we can cancel these terms and we obtain a well-defined explicit formula even when $\ga_1=1$ or $\ga_{d+1}=1$. Indeed, if we assume that at least one of the recurrence coefficients in the difference operator $F(\cL_{1,i},\cL_{i,d+1},\cL_{1,j},\cL_{j,d+1},\cL_{1,d+1})$ is not zero when, say, $\ga_1=1$, then the equation
$$F(\cL_{1,i},\cL_{i,d+1},\cL_{1,j},\cL_{j,d+1},\cL_{1,d+1})P_{\nu}(x;\ga)=0$$
will imply that the polynomials $P_{\nu}(x;\ga)$ satisfy a nontrivial recurrence relation in the indices $\nu$, with coefficients depending only on $\nu$. This means that the polynomials $P_{\nu}(x;\ga)$ are linearly dependent, which is impossible since they form a basis for the space of polynomials in the variables $x_1,\dots,x_d$, as long as the parameters $\ga=(\ga_1,\dots,\ga_{d+1})$ satisfy the conditions in equation \eqref{5.3}. 

\subsection{Irreducibility}\label{ss5.2}

Let 
$$\la_j(\nu)=-|\bnu^{j}|(|\bnu^{j}|+|\bga^{j}|+d+1-j)$$ 
denote the eigenvalue of the operator $\cM_j$ in equation \eqref{5.5}.
Using that fact that the parameters satisfy the conditions in equation \eqref{5.3}, we see that for $\nu,\mu\in\Nset_0^d$
\begin{equation}
\la_j(\nu)=\la_j(\mu) \text{ for all }j=1,\dots, d\qquad \text{ if and only if }\qquad \nu=\mu.
\end{equation}
Therefore, the spectral equations \eqref{5.3} fix the polynomials $P_{\nu}(x;\ga)$, up to unessential factors depending only on $\nu$.

We prove below that the $\fA_{d}(\ga)$-module $\cV^{d}_n(\ga)$ is irreducible, by induction on $d$, extending the constructions in Section~\ref{ss4.2}. Since we have already proved the statement in dimensions $2$ and $3$, we assume below that $d\geq 4$ and that the statement is true in dimension $d-1$.

First, we identify two subspaces of $\cV^{d}_n(\ga)$, which are irreducible modules over subalgebras of $\fA_{d}(\ga)$ isomorphic to $\fA_{2}$ and $\fA_{d-1}$, respectively.

For every $k\in\Nset_0$ such that $k\leq n$ we define the subspace
\begin{equation}\label{5.19}
\cW^{1}_{n,k}(\ga)=\Span\{P_{\nu}(x;\ga):\nu_1=k,\; \nu_2+\cdots+\nu_n=n-k\},
\end{equation}
of $\cV^{d}_n(\ga)$. We can identify this space with the module $\cV^{d-1}_n(\bga^{2})$ over the algebra $\fA_{d-1}(\bga^{2})$ generated by the operators $\cL_{i,j}$ for $2\leq i<j\leq d+1$. Indeed, when $2\leq i<j\leq d+1$ the operator $\cL_{i,j}$ contains no differentiation with respect to $x_1$, hence they will commute with the first term in the product on the right-hand side of \eqref{5.1}. Moreover,
if we introduce new variables $y_1,\dots,y_{d-1}$ via the formulas 
\begin{equation}\label{5.20}
y_j=\frac{x_{j+1}}{1-x_1}, \qquad\text{ for }j=1,\dots,d-1,
\end{equation}
then, up to a factor independent of $x_2,\dots,x_d$, the product of the last $d-1$ terms
$$\prod_{k=2}^d \left(1-|\bx_{k-1}| \right)^{\nu_k} 
               p_{\nu_k}^{(a_k,\ga_k)}\left (\frac{2x_k}{1-|\bx_{k-1}|} -1\right)$$
in \eqref{5.1} coincides with the polynomial $P_{\bnu^{2}}(y;\bga^{2})$. 
Furthermore, if we use \eqref{5.20} to change the variables in the operator 
$\cL_{i,j}$ for $2\leq i<j\leq d+1$, we obtain the operator $\cL_{i-1,j-1}$ for 
the $d-1$ sphere with parameters $\bga^{2}=(\ga_2,\dots,\ga_{d+1})$. Note also that if $\ga=(\ga_1,\dots,\ga_{d+1})$ satisfies the conditions \eqref{5.3}, then 
$\bga^{2}=(\ga_2,\dots,\ga_{d+1})$ also satisfies these conditions when $d$ is replaced by $d-1$. Therefore, by the induction hypothesis, the module $\cW^{1}_{n,k}(\ga)$ over the algebra  $\fA_{d-1}(\bga^{2})$ generated by the operators $\cL_{i,j}$ for $2\leq i<j\leq d+1$ is irreducible. 

The second subspace of $\cV^d_n(\ga)$ is 
\begin{equation}\label{5.21}
\cW^{0}_{n}(\ga)=\Span\{P_{\nu}(x;\ga):\nu_1+\nu_2=n,\;\nu_3=\cdots=\nu_d=0\}.
\end{equation}
Note that the polynomial $P_{\nu_1,\nu_2,0,\dots,0}(x;\ga)$ coincides with the two-variable Jacobi polynomial $P_{\nu_1,\nu_2}(x_1,x_2;\gat)$ in \eqref{3.2}, where $\gat_1=\ga_1$, $\gat_2=\ga_2$ and $\gat_3=|\bga^{3}|+d-2$. Moreover, since $\cW^{0}_{n}(\ga)$ contains polynomials depending only on $x_1$ and $x_2$, the operators $\pd_{x_j}$ will act as the zero operators for $j=3,\dots,d$ on this space. If we set 
$$\cLh_{1,3}=\sum_{j=3}^{d+1}\cL_{1,j}\qquad\text{ and }\qquad \cLh_{2,3}=\sum_{j=3}^{d+1}\cL_{2,j},$$
then using the above comments and the explicit formulas \eqref{2.5}-\eqref{2.6} we see that
\begin{align*}
\cLh_{1,3}\Big|_{\cW^{0}_{n}(\ga)}&=x_1(1-x_1-x_2)\pd_{x_1}^2+[(\ga_1+1)(1-x_1-x_2)-(|\bga^{3}|+d-1)x_1]\pd_{x_1},\\
\cLh_{2,3}\Big|_{\cW^{0}_{n}(\ga)}&=x_2(1-x_1-x_2)\pd_{x_2}^2+[(\ga_2+1)(1-x_1-x_2)-(|\bga^{3}|+d-1)x_2]\pd_{x_2}.
\end{align*}
If we denote by $\hat{\fA}$ the algebra generated by $\cLh_{1,2}=\cL_{1,2}$, $\cLh_{1,3}$ and $\cLh_{2,3}$, then from the above formulas it is clear that this algebra will preserve the space $\cW^{0}_{n}(\ga)$. Moreover, we can identify the $\hat{\fA}$-module $\cW^{0}_{n}(\ga)$ with the $\fA_2(\gat)$-module $\cV^2_n(\gat)$ constructed in Section~\ref{se3}. Finally,  if the parameters $\ga=(\ga_1,\dots,\ga_{d+1})$ satisfy \eqref{5.3}, then the parameters $\gat=(\gat_1,\gat_2,\gat_3)$ satisfy the conditions in \eqref{3.3}, and therefore this module is irreducible.

Now we can prove that the $\fA_{d}(\ga)$-module $\cV^{d}_n(\ga)$ is irreducible. Suppose that $\cV\neq\{0\}$ is an $\fA_{d}(\ga)$-submodule of $\cV^{d}_n(\ga)$. The operators $\cM_1,\dots,\cM_d$ can be simultaneously diagonalized on $\cV$ and therefore, there exists at least one polynomial $P_{\mu}(x;\ga)$ which belongs to $\cV$. 
Since $P_{\mu}(x;\ga)\in \cW^{1}_{n,\mu_1}(\ga)$ and $\cW^{1}_{n,\mu_1}(\ga)$ is an irreducible module over the algebra generated by $\cL_{i,j}$, for $2\leq i<j\leq d+1$, it follows that 
$\cW^{1}_{n,\mu_1}(\ga)\subset\cV$. In particular, this means that $P_{\mu_1,n-\mu_1,0,\dots,0}(x;\ga)\in\cV$. We use now that $P_{\mu_1,n-\mu_1,0,\dots,0}(x;\ga)\in \cW^{0}_{n}(\ga)$ and that $\cW^{0}_{n}(\ga)$ is an irreducible module over the algebra generated by the operators $\cLh_{1,2}$, $\cLh_{1,3}$, $\cLh_{2,3}$. This implies that $\cW^{0}_{n}(\ga)\subset\cV$. Pick arbitrary $\nu_1\in\Nset$, such that $\nu_1\leq n$. Then $P_{\nu_1,n-\nu_1,0,\dots,0}\in \cW^{0}_{n}(\ga)\subset\cV$. Using now that  $P_{\nu_1,n-\nu_1,0,\dots,0}(x;\ga)\in \cW^{1}_{n,\nu_1}(\ga)$ and that $\cW^{1}_{n,\nu_1}(\ga)$ is an irreducible module over the algebra generated by $\cL_{i,j}$, for $2\leq i<j\leq d+1$, we conclude that $\cW^{1}_{n,\nu_1}(\ga)\subset\cV$. Since this true for every $\nu_1\leq n$, it follows that $\cV^{d}_n(\ga)=\cV$, completing the proof of the irreducibility.

\end{document}